\documentclass[fleqn,10pt]{wlscirep}
\usepackage[utf8]{inputenc}
\usepackage[T1]{fontenc}
\usepackage{lineno}

\title{Near-real-time estimates of daily CO$_2$ emissions from 1500 cities worldwide}

\author[1,*]{Da Huo}
\author[1]{Xiaoting Huang}
\author[1]{Xinyu Dou}
\author[2]{Philippe Ciais}
\author[1]{Yun Li}
\author[1]{Zhu Deng}
\author[3]{Yilong Wang}
\author[1]{Duo Cui}
\author[4]{Fouzi Benkhelifa}
\author[1]{Taochun Sun}
\author[1]{Biqing Zhu}
\author[5]{Geoffrey Roest}
\author[5]{Kevin R. Gurney}
\author[1]{Piyu Ke}
\author[1]{Rui Guo}
\author[1]{Chenxi Lu}
\author[1]{Xiaojuan Lin}
\author[6]{Arminel Lovell}
\author[6]{Kyra Appleby}
\author[7]{Philip L. DeCola}
\author[8]{Steven J. Davis}
\author[1,*]{Zhu Liu}
\affil[1]{Department of Earth System Science, Tsinghua University, Beijing, 100084, China}
\affil[2]{Laboratoire des Sciences du Climate et de l'Environnement LSCE, Orme de Merisiers 91191 Gif-sur-Yvette, France}
\affil[3]{Key Laboratory of Land Surface Pattern and Simulation, Institute of Geographical Sciences and Natural Resources Research, Chinese Academy of Sciences, Beijing, 100101, China}
\affil[4]{Nexqt, City Climate Intelligence, France}
\affil[5]{School of Informatics, Computing, and Cyber Systems, Northern Arizona University, Flagstaff, AZ, 86011, USA}
\affil[6]{CDP Worldwide, London, UK}
\affil[7]{Department of Atmospheric and Oceanic Sciences, University of Maryland, College Park, MD, 20742, USA}
\affil[8]{Department of Earth System Science, University of California, Irvine, 3232 Croul Hall, Irvine, CA, 92697-3100, USA}

\affil[*]{corresponding authors: Da Huo (dh2107@tsinghua.edu.cn), Zhu Liu (zhuliu@tsinghua.edu.cn)}

\begin{abstract}
Building on near-real-time and spatially explicit estimates of daily carbon dioxide (CO$_2$) emissions, here we present and analyze a new city-level dataset of fossil fuel and cement emissions. Carbon Monitor Cities provides daily, city-level estimates of emissions from January 2019 through December 2021 for 1500 cities in 46 countries, and disaggregates five sectors: power generation, residential (buildings), industry, ground transportation, and aviation. The goal of this dataset is to improve the timeliness and temporal resolution of city-level emission inventories and includes estimates for both functional urban areas and city administrative areas that are consistent with global and regional totals. Comparisons with other datasets (i.e. CEADs, MEIC, Vulcan, and CDP) were performed, and we estimate the overall uncertainty range to be ±21.7\%. Carbon Monitor Cities is a near-real-time, city-level emission dataset that includes cities around the world, including the first estimates for many cities in low-income countries. A more complete description of this dataset is published in Scientific Data \hyperlink{https://doi.org/10.1038/s41597-022-01657-z}{https://doi.org/10.1038/s41597-022-01657-z}. 

\end{abstract}
\begin{document}

\flushbottom
\maketitle

\thispagestyle{empty}
\section*{Background \& Summary}

More than 60\% of global fossil-fuel CO$_2$ emissions are produced in cities \cite{Duren2012, Seto2014}, and high-quality city-level emissions inventories are urgently needed to support international climate mitigation efforts \cite{IPCC2014urban, Gurney2020, Gurney2021}. For example, many cities have adopted goals of reaching net-zero emissions by 2030 or 2050, which require them to monitor and report emissions on a timely basis \cite{Seto2021}. Unfortunately, a global, open, and harmonized dataset of city-level emission inventories is yet lacking \cite{Kona2018,Nangini2019}. Instead, most CO$_2$ emission inventories are conducted at the country level, as city-level fossil fuel consumption data are more difficult to acquire \cite{Chen2021}. Furthermore, many inventories–including national inventories reported to the United Nations Framework Convention on Climate Change (UNFCCC) often lag reality by one years or more \cite{Liu2020}. Thus, many city-level mitigation efforts are hampered by a lack of timely and high-quality emissions data with which to set benchmarks and monitor progress \cite{Ramaswami2021, Bulkeley2010, DAvignon2010, Tan2017}.

City-level CO$_2$ emissions may refer to either the CO$_2$ emissions produced within the territory of a city or emissions related to all the goods and services consumed in a city, which often include substantial emissions produced outside the city boundary \cite{Ramaswami2011, Chen2019, Nangini2019}. The in-boundary emissions are typically referred to as scope 1, emissions from imported electricity as scope 2, and all other trans-boundary emissions associated with other city activities are referred to as scope 3 \cite{Chen2019, Wiedmann2021}. Three conventional approaches have been used to attribute CO$_2$ emissions to cities: purely geographic production-based accounting, community infrastructure-based accounting (geographic-plus), and consumption-based accounting \cite{Ramaswami2011, Chen2019}. These approaches can provide good estimates for major cities that disclose high-quality energy consumption data \cite{Ramaswami2011}. However, they can not be readily applied to a larger scale, where city-specific data are largely absent, especially for smaller cities \cite{Moran2021}. Downscaling represents a solution to the scalability issue. Some recent studies use economic input-output (IO) tables down-scaled from national statistics to attribute emissions to cities \cite{Jing2018, Wiedmann2021}, and other studies use spatial proxies to disaggregate national or sub-national emissions to finer scales. Popular spatial proxies include night-time light imagery and existing gridded emission maps, such as the Emission Database for Global Atmospheric Research (EDGAR) \cite{Marcotullio2014, Crippa2019, Crippa2020}, which also relies on other spatial data like population density and road networks. Downscaling has been used to construct multiple city-level datasets that cover a large number of cities \cite{Marcotullio2014, Chen2019}, and we adopt a similar approach in this study.

Cities that disclose their emissions typically follow protocols or standards such as the Global Protocol for Community-Scale Greenhouse Gas Emission Inventories (GPC) \cite{Fong2015}, the International Council for Local Environmental Initiatives (ICLEI) \cite{Kona2021}  or the ICLEI-USA \cite{Ramaswami2011}. However, the reliability of these self-reported inventories is difficult to assess due to the lack of peer-review \cite{Nangini2019, Long2021, Gurney2021}. Inter-dataset comparison is also difficult due to inconsistent definitions of spatial and temporal scales, protocols, sector coverage, activity data sources, and accounting methods \cite{Kennedy2009, Kennedy2010, Nangini2019, Gurney2021, Kona2021}, and many original input data are untraceable \cite{Nangini2019}. Therefore, a methodological framework that supports inter-dataset comparisons and calibration is yet to be developed for cities.

Where city-level emissions inventories exist, they often rely on data provided by organizations such as the China Emission Accounts and Datasets (CEADs: \hyperlink{https://www.ceads.net/}{https://www.ceads.net/} \cite{Shan2020}), the Multi-resolution Emission Inventory (MEIC, \hyperlink{http://meicmodel.org/}{http://meicmodel.org/} \cite{Zheng2014,Liu2015}), and the Carbon$n$ Climate Registry (\hyperlink{https://carbonn.org/}{https://carbonn.org/}) \cite{Mi2016, Nangini2019}, or inventory warehouses like the CDP worldwide (CDP: \hyperlink{https://www.cdp.net/}{https://www.cdp.net/}). But the coverage, timeliness, and temporal resolution of these data are not always sufficient to support agile and informed decision making. For example, although several high-quality datasets are available for high-income countries, such as the Covenant of Mayors database (\hyperlink{https://www.globalcovenantofmayors.org/}{https://www.globalcovenantofmayors.org/} \cite{Croci2017, Kona2021}) and OpenGHGmap (\hyperlink{https://openghgmap.net/}{https://openghgmap.net/} \cite{Moran2021}) for EU cities, and the Vulcan and Hestia datasets for the US cities (\hyperlink{https://vulcan.rc.nau.edu/}{https://vulcan.rc.nau.edu/},   \hyperlink{https://hestia.rc.nau.edu/}{https://hestia.rc.nau.edu/} \cite{Gurney2019, Gurney2020, Gurney2021}), the tension between human development and decarbonization requires an increasing focus on rapidly expanding cities in low-income and emerging regions in South America, South and Southeast Asia, Africa and the Middle East where high-quality emission inventories are lacking \cite{Andrew2020, WangRR2021}. Moreover, most existing city-level inventories have the issues of long time lag and low temporal resolution. Recently, the methodological frameworks for estimating near-real-time (NRT) daily CO$_2$ emissions have been developed and successfully used for studying the impacts of COVID-19 on global CO$_2$ emissions \cite{Liu2020b, LeQuere2020}. Here, we build on these approaches to provide NRT daily emission estimates for hundreds of cities worldwide, including many in low-income regions.

The dataset presented in this paper also provides a possible solution to address the inconsistency between administrative emissions versus community-wide emissions. Differences in spatial scope and accounting methods inevitably complicate comparisons among cities \cite{Chen2019, Nangini2019}, and one possible solution is to compile the inventories based on different functional zones of the city, such as differentiating the core and commuting zones  \cite{Kona2021}. We provide a global harmonized dataset that consistently quantifies production-based emissions from core administrative areas in top emitting countries and, separately, emissions from the world's major metropolitan/functional urban areas.  

\section*{Methods}
\subsection*{Workflow}
\begin{figure}[ht]
\centering
\includegraphics[width=\linewidth]{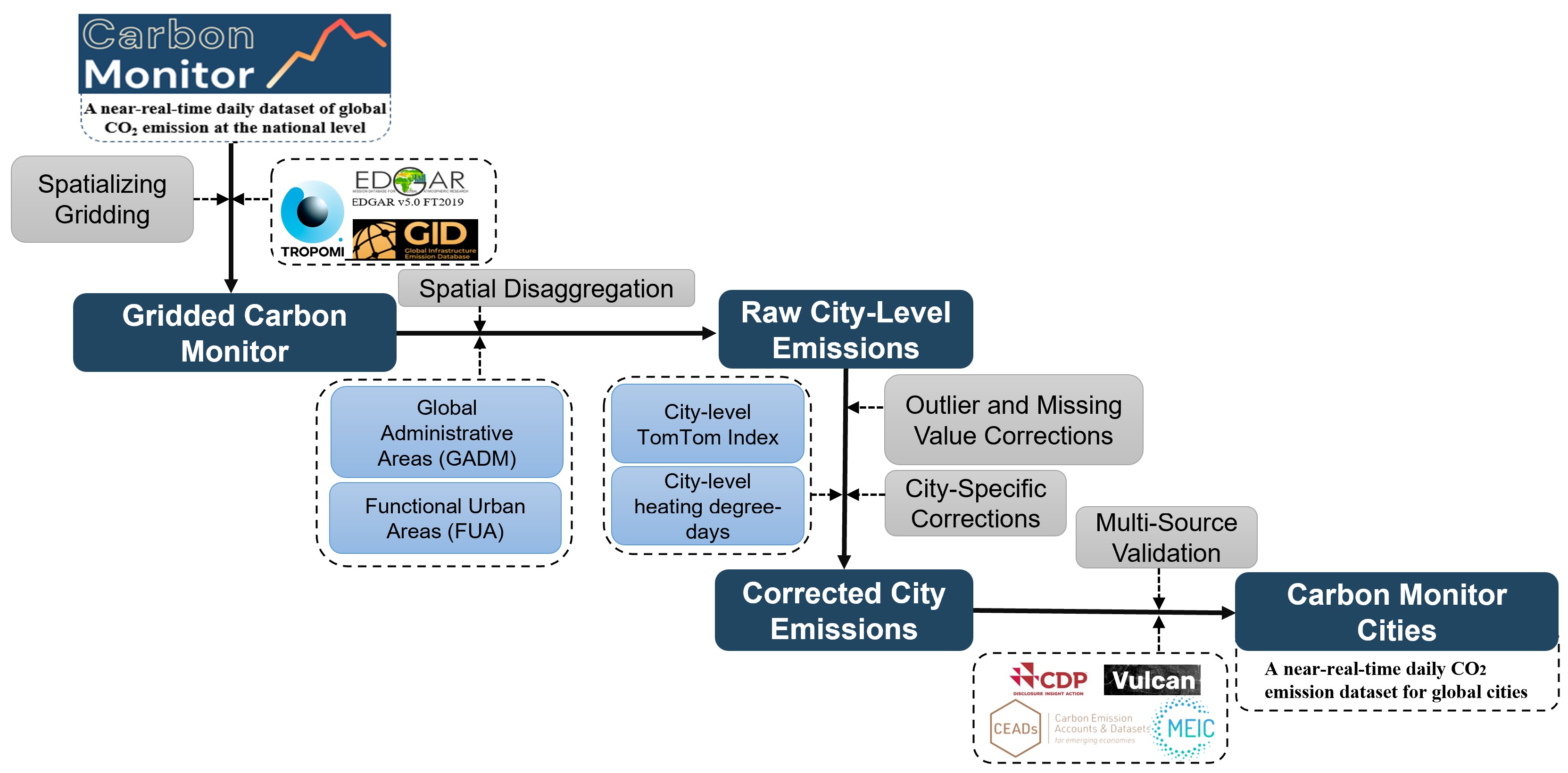}
\caption{Flowchart illustrates the main workflow and data used in each stage.}
\label{fig:fig1}
\end{figure}

Carbon Monitor Cities (CM-Cities) is downscaled from the Carbon Monitor, which is a NRT national level emission dataset at a global scale \cite{Liu2020}. Specifically, CM-Cities is produced following a four-stage workflow (Fig. \ref{fig:fig1}). The first stage mainly involves the construction of Global Gridded Daily CO$_2$ Emission Datasets (GRACED) \cite{Dou2022}, which are daily emission maps generated by spatializing Carbon Monitor daily emissions using the Global Carbon Grid (GID), the Emissions Database for Global Atmospheric Research (EDGAR) and TROPOspheric Monitoring Instrument (TROPOMI). GRACED covers seven sectors (power, industry, residential and commercial buildings, ground transportation, domestic aviation, international aviation, and international shipping) and provides NRT emission maps for fossil fuel combustion and cement production with a global spatial resolution of 0.1$^{\circ}$ by 0.1$^{\circ}$ and a temporal resolution of one day. GRACED is an intermediate gridded dataset between the Carbon Monitor and CM-Cities, and the methods for generating this gridded dataset is described in a later section.

In the second stage, we disaggregated the gridded daily emissions into cities based on two types of city areas : Global Administrative Areas (GADM) and Functional Urban Areas (FUA) to address the definition differences of 'a city' in different countries. The FUA is defined by the Organisation for Economic Co-operation and Development (OECD) and the European Union as the high-density urban centres plus their surrounding commuting zones \cite{eFUA2021}. For OECD countries, we used the OECD FUA, which provides higher quality FUA for OECD countries (\hyperlink{https://www.oecd.org/regional/regional-statistics/functional-urban-areas.htm}{https://www.oecd.org/regional/regional-statistics/functional-urban-areas.htm} \cite{FUA}). For other countries, the Global Human Settlement FUA is used (\hyperlink{https://ghsl.jrc.ec.europa.eu/ghs\_fua.php}{https://ghsl.jrc.ec.europa.eu/ghs\_fua.php} \cite{eFUA2019}). The GADM level-2 is used for prefecture-level cities in China and counties in the United States. The details of the features and the usage of  FUA and GADM datasets are described in later sections. The spatial downscaling/disaggregation is performed by first converting the FUA and GADM shapefiles into raster datasets with a unique ID assigned to each city. Then the raster city area maps are used as masks to extract the matching grid cells in the GRACED emission maps. We then aggregate emission values for grid cells that correspond to the same city mask to yield the total sectoral emission value for a given city.

In the third stage, we use city-level data to correct for the residential and ground transport sectors to address the bias in raw city-level inventories from the second stage. We use city-specific TomTom congestion data and daily heating degree days (HDD) for the corrections.

The fourth stage involves error correction and data validation. We first identify and remove outliers (which are mostly errors introduced by previous processing steps and/or from the source data) using statistical approaches. We then collected city-level inventories from other datasets (mostly annual data) and compared them to our results for validation. The detailed procedures for these processes are described in later sections.

CM-Cities includes city-level emission inventories from 01/01/2019 to 31/12/2021 for five main sectors: 1. power generation, 2. residential and commercial buildings, 3. industrial production, 4. ground transportation, and 5. aviation. These five sectors combined account for over 70\% of fossil fuel CO$_2$ emissions from a city \cite{Gurney2019}. 

\subsection*{Coverage}
\begin{figure}[ht]
\centering
\includegraphics[width=\linewidth]{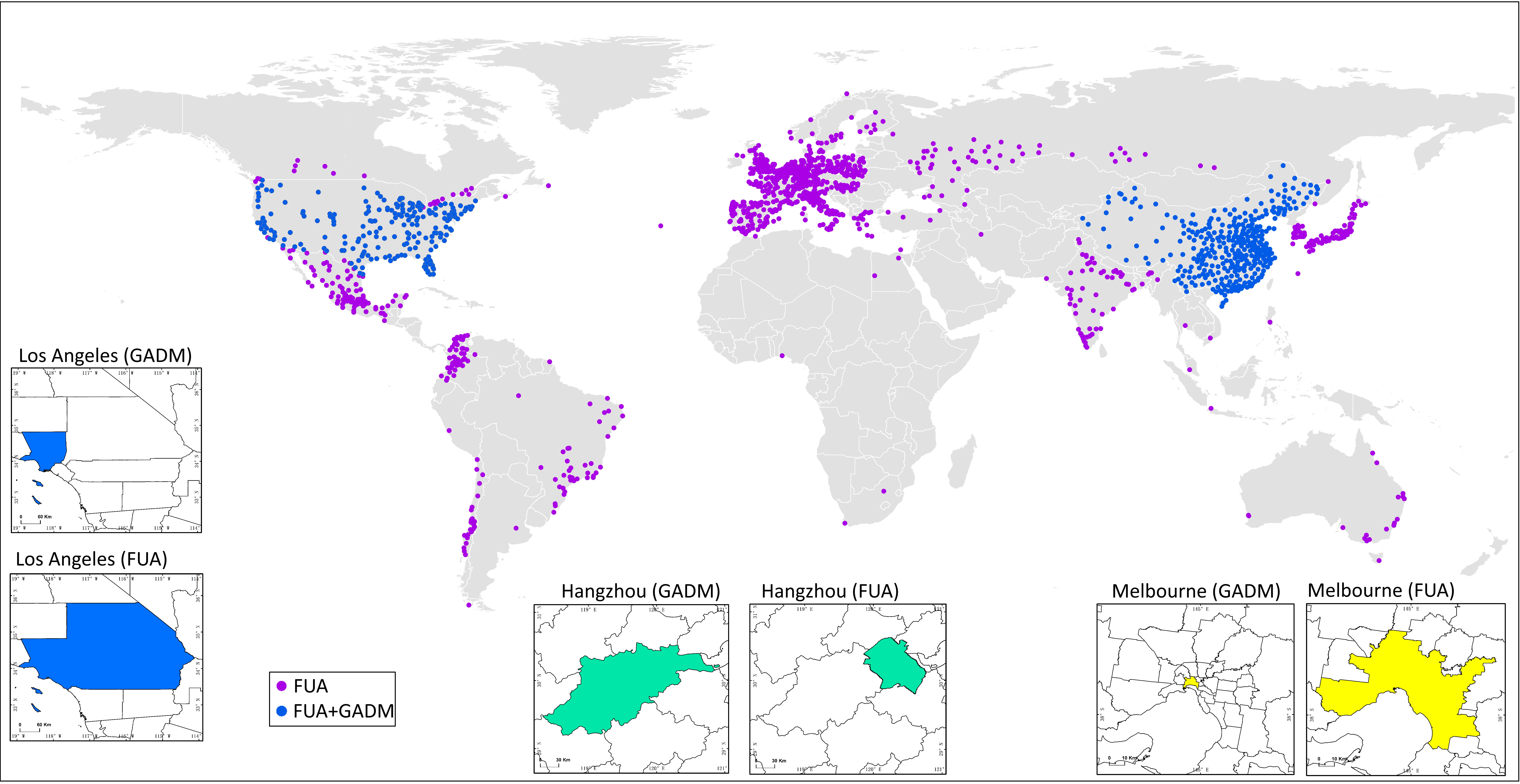}
\caption{Map showing all the cities covered in this dataset. Purple dots indicate cities with emissions estimated based on functional urban areas (FUA), and blue dots indicate cities with emissions estimated based on both FUA and administrative areas (GADM). Subplots depict examples of the comparison between the administrative city area versus the functional urban areas for cities in different regions.}
\label{fig:fig2}
\end{figure}

CM-Cities covers 1500 cities in 46 counties (Fig. \ref{fig:fig2}). Most of the cities are clustered in Europe, Asia, North and South America. Major cities in Oceania and Africa are also included. Figure \ref{fig:fig2} also shows comparisons between the FUA and the GADM for Los Angeles (US) , Hangzhou (China), and Melbourne (Australia). The FUA typically covers a larger area than the administrative area, but for cities in some countries, such as China, the FUA is typically smaller than the administrative city area. The use of both area definitions facilitates dataset comparisons, which is highlighted for cities in China and the United States. These two different spatial scopes also provide critical information for differentiating administrative emissions versus community-wide emissions.

\subsection*{Near-Real-Time Daily Emissions by Sector}
CM Cities is downscaled from the GRACED dataset, in which spatial distribution and daily variations of emissions are combined. This section describes the methods for estimating NRT daily emissions from a temporal perspective, and the next section describes the spatial gridding procedure. The estimation of daily emission variations follows the Carbon Monitor national dataset \cite{Liu2020, Liu2020b}, which provides daily fossil fuel CO$_2$ emissions since January 1st, 2019 on the global and national levels, with detailed estimates in 7 main sectors, i.e., power, industry, ground transport, residential (including commercial), domestic aviation, international aviation, and international shipping . Emissions from international bunkers (including the international aviation sector and international shipping sector) are only accounted for at the global level and usually excluded from the national territorial emissions according to the IPCC guidelines. Thus, CM-Cites consider the other 5 sectors, i.e., power, industry, ground transport, residential, and domestic aviation. 

\subsubsection*{Power Sector}
Daily power generation data are acquired from multiple open data sources depending on the country (Table \ref{tab:tab1}), which provides live power generation data with a daily or hourly resolution, and accounts for more than 70\% of the total CO$_2$ emissions in the power sector \cite{Liu2020}. The emission factors are estimated using EDGAR's electricity emissions, divided by our collection of coal-fired electricity data in various countries. The daily emissions are estimated as:

\begin{equation}
Emis_{power,daily}=Emis_{power,yearly}(AD_{power,daily}/AD_{power,yearly}),
\end{equation}

\noindent where $AD$ is the power generation. For emissions from other countries (countries not listed in Table \ref{tab:tab1}), we assumed a linear relationship between daily global emission and daily total emissions from these countries, and then adjusted the emissions for countries that adopted lock-down measures during the COVID-19 following the method used by the Carbon Monitor national dataset \cite{Liu2020}.

\subsubsection*{Industry Sector}
For the industry sector, the daily emissions are calculated from the monthly industrial production index and the daily power generation data. Monthly industrial production data are acquired from several datasets (Table \ref{tab:tab2}). The monthly CO$_2$ emissions estimated from the Industrial Production Index (IPI) are then disaggregated into a daily scale using daily power generation data. This approach is based on two assumptions: 1. A linear relationship exists between daily industrial production and industrial fossil fuel use. 2. A linear relationship exists between daily industry activity and daily electricity production \cite{Liu2020}.The monthly and daily industry emissions are estimated following:

\begin{equation}
Emis_{ind,monthly,currentyear,c}=Emis_{ind,yearly,2019,c}(IPI_{monthly,currentyear,c}/IPI_{yearly,2019,c}),
\end{equation}

\begin{equation}
Emis_{ind,daily}=Emis_{ind,monthly}(Elec_{daily}/Elec_{monthly}),
\end{equation}

\noindent where $Emis_{ind,monthly,currentyear,c}$ is the monthly industry emissions for country $c$ in current year, $Emis_{ind,yearly,2019,c}$ is the yearly industry emissions for country $c$ in 2019 (year of the latest update of baseline emissions), $IPI$ is the corresponding Industrial Production Index. $Emis_{ind,daily}$ and $Emis_{ind,monthly}$ are the daily and monthly industry emissions, respectively. $Elec_{daily}$ and $Elec_{monthly}$ are the daily and monthly electricity production, respectively. For countries not listed in Table \ref{tab:tab2}, the industry sector emissions are estimated in the same way as for the power sector.

\subsubsection*{Ground Transport Sector }
\begin{figure}[ht]
\centering
\includegraphics[width=12cm]{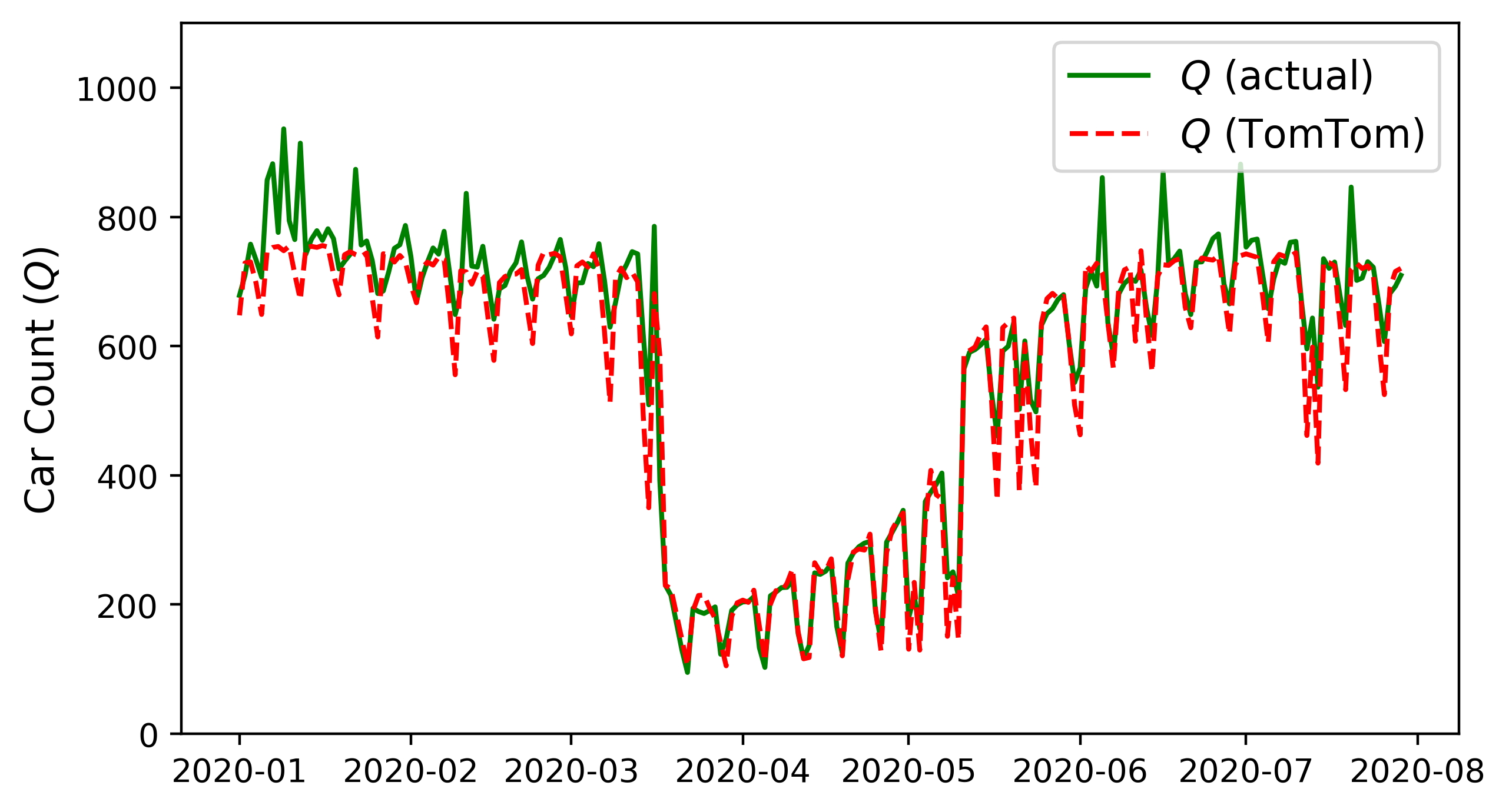}
\caption{Comparison between the actual and TomTom estimated hourly car counts on the measured roads in Paris. TomTom-based estimates accurately depicted the drop in traffic during the lock down period in 2020.}
\label{fig:fig3}
\end{figure}

Daily emissions from ground transportation are estimated using TomTom live congestion index and EDGAR road transportation emissions. The TomTom traffic congestion level represents the extra time spent on a trip in congested conditions, as a percentage, compared to uncongested conditions. TomTom congestion level data were obtained for more than 400 cities around the world at a temporal resolution of one hour. This approach permits the estimation of NRT emissions from ground transportation with a temporal resolution up to one hour. The TomTom live congestion level data was proven to be highly accurate for most cities, and Carbon Monitor has successfully adopted this approach \cite{Liu2020}. Note that a zero-congestion level means the traffic is fluid or ‘normal’ but does not mean there are no vehicles and zero emissions. The lower threshold of emissions when the congestion level is zero was estimated using real-time data from an average of 60 roads in the city of Paris \cite{Wang2021}. TomTom data accurately depicts the traffic volume using a sigmoid function-based regression (Eq. \ref{eq:trans1}), and Figure \ref{fig:fig3} is a comparison between the actual and TomTom estimated hourly car counts on the measured roads in Paris. The estimated traffic volume is then used to allocate the EDGAR on-road emissions to each day (Eq. \ref{eq:trans2}). 

\begin{equation}\label{eq:trans1}
Q_{d}=a+\frac{\beta X^{\gamma}}{{\lambda}^{\gamma}+X^{\gamma}},
\end{equation}

\begin{equation}\label{eq:trans2}
Emis_{trans,d} = Q_{d}\frac{Emis_{onroad}}{\sum_{d=1}^{n}Q_{d}},
\end{equation}

\noindent where $Q_{d}$ is the mean vehicle number per hour in day $d$, $X$ is the daily mean TomTom congestion level data, and $a$, $\beta$, $\gamma$, $\lambda$ are regression parameters, $Emis_{trans,c,d}$ is the ground transport emissions in day $d$, $Emis_{onroad}$ is the annual EDGAR road transportation emissions, $n$ is the number of days in a year. For cities not covered by TomTom, we assumed that the emission changes follow the mean changes of other cities in the country. If no city in the country has TomTom data, then the relative emission changes are assumed to follow the same pattern of the total emissions from all TomTom-covered countries.

\subsubsection*{Residential Sector}
Carbon Monitor uses the fluctuation of air temperature to capture the daily variations in the energy consumption of residential and commercial buildings. The assumption associated with this method is that the heating demand, which is the largest contribution to the daily variability in emissions for this sector, is strongly governed by air temperature \cite{Spoladore2016}, which determines the HDD (cooling in summer mainly consumes electricity that is covered in the power sector). This approach uses population-weighted HDD for different geographic locations for each day based on the ERA-5 reanalysis of air temperature \cite{ERA5} and also accounts for temperature-independent cooking emissions following EDGAR. The EDGAR residential emissions are then downscaled to daily values based on daily variations in population-weighted heating degree days. 

\begin{equation}
Emis_{res,c,d}=Emis_{res,c,m}R_{heating,c,m}\frac{HDD_{c,d}}{\sum_{m}HDD_{c,d}}+Emis_{res,c,m}\frac{1-R_{heating,c,m}}{N_{m}},
\end{equation}

\begin{equation}\label{eq:hdd}
HDD_{c,d}=\sum_{g}R_{pop,g}He(18-T_{g,d}),
\end{equation}

\noindent where $Emis_{res,c,d}$ and $Emis_{res,c,m}$ are the residential emissions for country $c$ in day $d$ and month $m$ respectively, $R_{heating,c,m}$ is the percentage of residential emissions from heating demand in country $c$ in month $m$, $HDD_{c,d}$ is the population-weighted heating degree day for country $c$ in day $d$,  $N_m$ is the number of days in month $m$, $R_{pop,g}$ is the ratio of the population in grid $g$ to the total national population, which is acquired from the Gridded Population of the World, version 4 \cite{Doxsey2015}, $He$ is a Heaviside step function that converts any negative values to zero, $T_{g,d}$ is the average air temperature in Celsius for grid $g$ in day $d$ at 2 meters derived from ERA5 \cite{ERA5}, and 18 is a HDD reference temperature of 18°C.

\subsubsection*{Aviation Sector}
Emissions in the aviation sector are computed from individual commercial flights data from the Flightradar24 database (\hyperlink{https://www.flightradar24.com}{https://www.flightradar24.com}). This sector covers domestic flights, and all airports around a city were selected even if they are not part of the FUA, but some airports are not covered by the GADM. The daily CO$_2$ emissions were estimated as the product of distance flown and a constant emission factor ($EF_{avi}$).

\begin{equation}
Emis_{avi}=DF\cdot EF_{avi},
\end{equation}

\noindent where $DF$ is distance flown, which is computed using great circle distance between the take-off, cruising, descent, and landing points for each flight and are cumulated over all flights. The emission factor per kilometer flown is assumed to be a constant for the mix of all aircraft from an airport (including regional, narrowbody passenger, widebody passenger, and freight operations) as the share of flight types has not significantly changed since 2019.

\subsection*{Gridded Daily CO$_2$ Emissions}
Carbon Monitor Cities disaggregates the Carbon Monitor national emissions to cities using the GRACED dataset developed by the Carbon Monitor team \cite{Dou2022}, which consists of emission maps generated by spatializing and gridding the daily national emission inventories from Carbon Monitor into grid cells. This was achieved by estimating spatial distribution proxies from satellite data and existing gridded products while maintaining consistency between ‘bottom-up’ accounting results and the spatial sum of the gridded results. Three datasets were used in producing GRACED: 1. The Global Carbon Grid (GID), which provides global CO$_2$ emissions data from major industry and power plant point sources with a resolution of 0.1$^{\circ}$ in 2019, 2. The Emissions Database for Global Atmospheric Research (EDGAR), which provides sectoral emissions as specified by the IPCC guidelines. 3. The NO$_2$ thermal chemical vapor deposition retrieval product from the Tropospheric Monitoring Instrument (TROPOMI) onboard the Sentinel‐5 Precursor satellite. Given that GID has higher data quality in fine-grained spatial scales and point sources of industries and power plants, the GID-based point sources and the EDGAR emission maps were combined for constructing GRACED (Eq. \ref{eq:grid1}). While the spatial emission patterns derived from GID and EDGAR (with latest updates in 2019) cannot accurately reflect the situation in 2020 and 2021, the NRT TROPOMI NO$_2$ retrievals were used as a proxy for CO$_2$ to capture the daily variability in CO$_2$ emission following GRACED \cite{Dou2022}. After several data processing steps, such as rolling-average and thresholding, the NO$_2$ data can reasonably indicate the spatial distribution of CO$_2$ sources \cite{Berezin2013}. Table \ref{tab:tab3} lists the gridded data used for producing GRACED. For the aviation sector, EDGAR's monthly data are used for spatial distribution (Eq. \ref{eq:grid2}). Thus, the gridded emissions $EmiGrid_{g,d,s}$ for grid $g$, date $d$ and sector $s$ were estimated as:

\begin{equation}\label{eq:grid1}
EmiGrid_{g,d,s}=CM_{c,d,s}\frac{GID_{g,s}}{\sum_{i=1}^{n} GID_{i,s}}\frac{EDGAR_{g,m,s}}{\sum_{j=1}^{12}EDGAR_{g,j,s}}\cdot12,
\end{equation}

\begin{equation}\label{eq:grid2}
EmiGrid_{g,d,avi}=CM_{c,d,avi}\frac{EDGAR_{g,m,avi}}{\sum_{i=1}^{n}EDGAR_{i,m,avi}},
\end{equation}

\noindent where $CM_{c,d,s}$ represents the value of Carbon Monitor national emission for country $c$, day $d$ and sector $s$. $s$ includes the power, industry, residential, and ground transport sectors. $avi$ is the aviation sector. $GID_{g,s}$ is the value of GID gridded CO$_2$ emissions for grid $g$ and sector $s$. $n$ is the total number of grids within this country and $j$ is the total number of month. $EDGAR_{g,m,s}$ represents the EDGAR gridded CO$_2$ emissions for grid $g$, sector $s$, and month $m$ which date $d$ belongs to.

\subsection*{City-level Spatial Disaggregation}
The spatial disaggregation is performed by first converting the city area shapefiles (FUA or GADM) into raster datasets with a unique ID assigned to each city. Then the raster city area maps are used as masks to extract the matching grid cells in GRACED emission maps. We then aggregate emission values for grid cells that correspond to the same city mask to yield the total sectoral emission value for a given city. For the aviation sector, emissions from all planes within the city's territory are included. The international shipping sector is not included in this dataset because most of the emissions from this sector occurred in the open ocean that cannot be allocated to specific cities. The jurisdiction issue also applies to the aviation sector, but we keep the territorial-based allocation approach in the dataset for completeness.
 
We use both the administrative areas and FUA because boundary definition has always been a problem in city-level inventory completion \cite{Chen2019}, as the administrative city areas in most countries do not reflect emissions from the larger commuting zones of a city, which may constitute a large part of the emissions, meanwhile, FUA represents the most viable spatial dataset for covering the more complete urban areas. In addition, FUA is clearly-defined and produced using a consistent method for cities worldwide, while the definition of administrative city areas may vary significantly across different countries. Therefore, this use of both spatial scopes represents a potential solution to differentiate administrative emissions versus community-wide emissions and makes inter-dataset comparisons easier as demonstrated in the validation section.

\subsection*{City-level Corrections}
The disaggregation from EDGAR spatial distributions is insufficient especially for the residential and ground transport sectors, because EDGAR uses a disaggregation of national sectoral totals per population for residential, and per road network for ground transport, which introduces bias to cities. Therefore, we correct these two sectors at individual city level. The ground transport sector emission is corrected using city-specific TomTom data (by applying Eqs. \ref{eq:trans1},\ref{eq:trans2} at city scale) for 416 cities worldwide that have their own NRT TomTom indices (list of these cities can be found in the documentation on Carbon Monitor website), which represent more accurate ground transport emission estimates for these cities. For cities that do not have their own TomTom data, we spatially disaggregate the national mean estimates following Eqs. \ref{eq:trans1},\ref{eq:trans2},\ref{eq:grid1}.
 
The residential sector is corrected using city-level HDD to overcome the bias of downscaling from national inventory. Specifically, we first calculate the daily mean HDD for each city from the  population-weighted HDD grid (Eq. \ref{eq:hdd}), and then use it as the baseline to compute a correction factor for each city by comparing it with the mean national HDD to update the emissions for the residential sector:

\begin{equation}
Emis_{res,i}=Emis0_{res,i}\left(1-\frac{HDD_{c}-HDD_{i}}{max(HDD_{c})}\right),
\end{equation}

\noindent where $Emis_{res,i}$ and $Emis0_{res,i}$ represent the corrected and uncorrected residential emissions, respectively for city $i$, $HDD_{c}$ is the mean daily HDD for the country, and $HDD_{i}$ is the mean daily HDD for city $i$.

\subsection*{Outlier Correction}
Outliers exist in the data mainly due to errors in the source datasets, such as mistakes in unit conversions or data entry, etc. To correct these outliers, we apply a statistical method based on intrinsic properties of the distribution of the emissions in the database. This allows more accurate identification of outliers that are likely to be the results of incorrect data entry. Similar statistical approaches have been successfully applied to correct for outliers in emission datasets \cite{Kona2021}. The outlier identification method is based on standard deviation (STD). Specifically, we consider an emission value as an outlier if the differences between the current value and its daily neighbours are both greater than 3 times the yearly STD for that sector (Eq. \ref{eq:std}). This threshold is determined by experimenting with data with known error and data with periodical high variation (e.g., weekday versus weekends for the ground transport emissions). These experiments determined the lower and upper bounds of the threshold such that it correctly identifies outliers and keeps the inherent variance within the data.

\begin{equation}\label{eq:std}
|x-x_{n+1}|>3\cdot STD\;and\;|x-x_{n-1}|>3\cdot STD.
\end{equation}

\subsection*{Limitations}
This dataset focuses on improving the timeliness, temporal resolution, and coverage of city-level inventories for studying NRT emission dynamics and also providing emission estimates for many cities in low-income regions. This dataset is derived from the gridded Carbon Monitor which is based on downscaled national inventories, combined with point sources and spatial distributions from GID and EDGAR, therefore, one limitation is the lack of using city-specific bottom-up activity data except for the ground transport sector, which may introduce additional uncertainties. We also noted that some input data may contain inherent errors and missing values (other than the above-mentioned outliers and discontinuities), especially for cities in less developed nations, we do not intend to fix this kind of errors in the source data without enough background information of the specific city, but we consider our results represent a meaningful first-order estimate for many of these cities that are lacking any emission inventories. Estimating NRT daily emissions is a relatively new research direction and require ongoing efforts to calibrate and update the workflow to improve data quality in the future.

\section*{Data Records}
CM-Cities provides NRT city-level emission inventories from 01/01/2019 to 31/12/2021 for 1500 cities in 46 counties. All data have gone through a validation process, in which we estimated the uncertainties and corrected errors. The attributes of the final dataset are listed in Table \ref{tab:tab4}, and the emission data are organized into spreadsheets. The definitions for sectors are consistent with the Carbon Monitor national inventories. Brief descriptions of the methods, sectors, coverage and uncertainty are also provided in Table \ref{tab:tab5}. Latest updates for selected cities and related information are available for view and download on our website \hyperlink{https://cities.carbonmonitor.org}{https://cities.carbonmonitor.org}. At the time of writing this article, this dataset has been updated to December 31, 2021 and the full dataset can be downloaded at Figshare \cite{Huo2022}. Future updates will also be available on our website.

\begin{itemize}
\item The file that contains functional urban area results for all cities (carbon-monitor-cities-all-cities-FUA.csv) has 8,114,886 data records (some cities have missing values). Separate data files are also provided for each of the 46 countries (carbon-monitor-cities-'CountryName'.csv).
\item The file that contains all administrative area results for Chinese cities (carbon-monitor-cities-China-GADM-prefecture.csv) has 1,885,120 data records, including 344 prefecture-level cities, and each city has 5480 data records.
\item The file that contains U.S. county-level results (carbon-monitor-cities-US-Counties.csv) has 1,720,720 data records, including 314 counties, and each county has 5480 data records.
\end{itemize}

\subsection*{Data Examples}
\begin{figure}[ht]
\centering
\includegraphics[width=\linewidth]{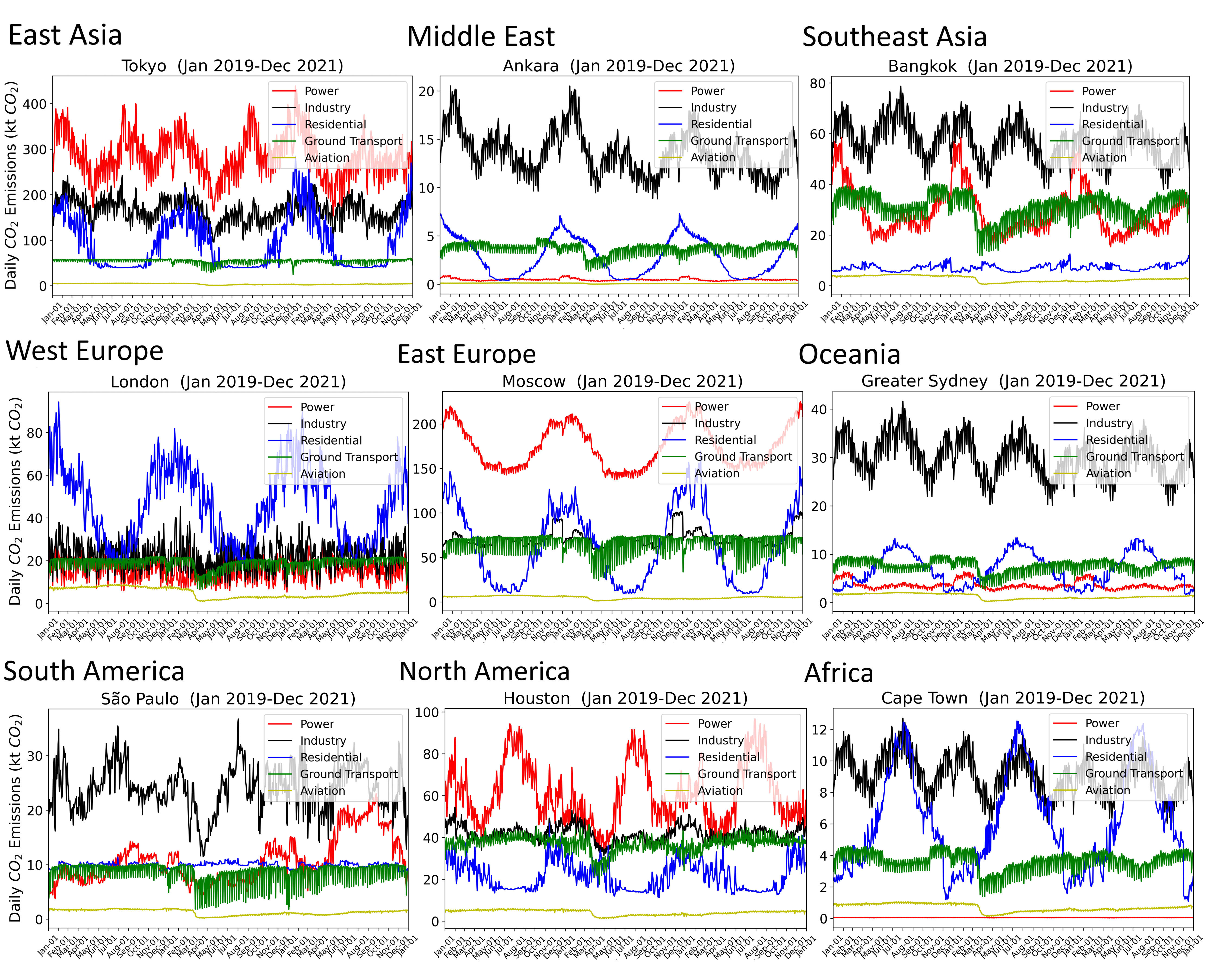}
\caption{Daily by sector CO$_2$ emissions for cities (FUA) in different regions of the world. Including Tokyo in East Asia, Ankara in the Middle East, Bangkok in Southeast Asia, London in West Europe, Moscow in East Europe, Greater Sydney in Oceania, São Paulo in South America, Houston in North America, and Cape Town in Africa.}
\label{fig:fig4}
\end{figure}

\begin{figure}[ht]
\centering
\includegraphics[width=12cm]{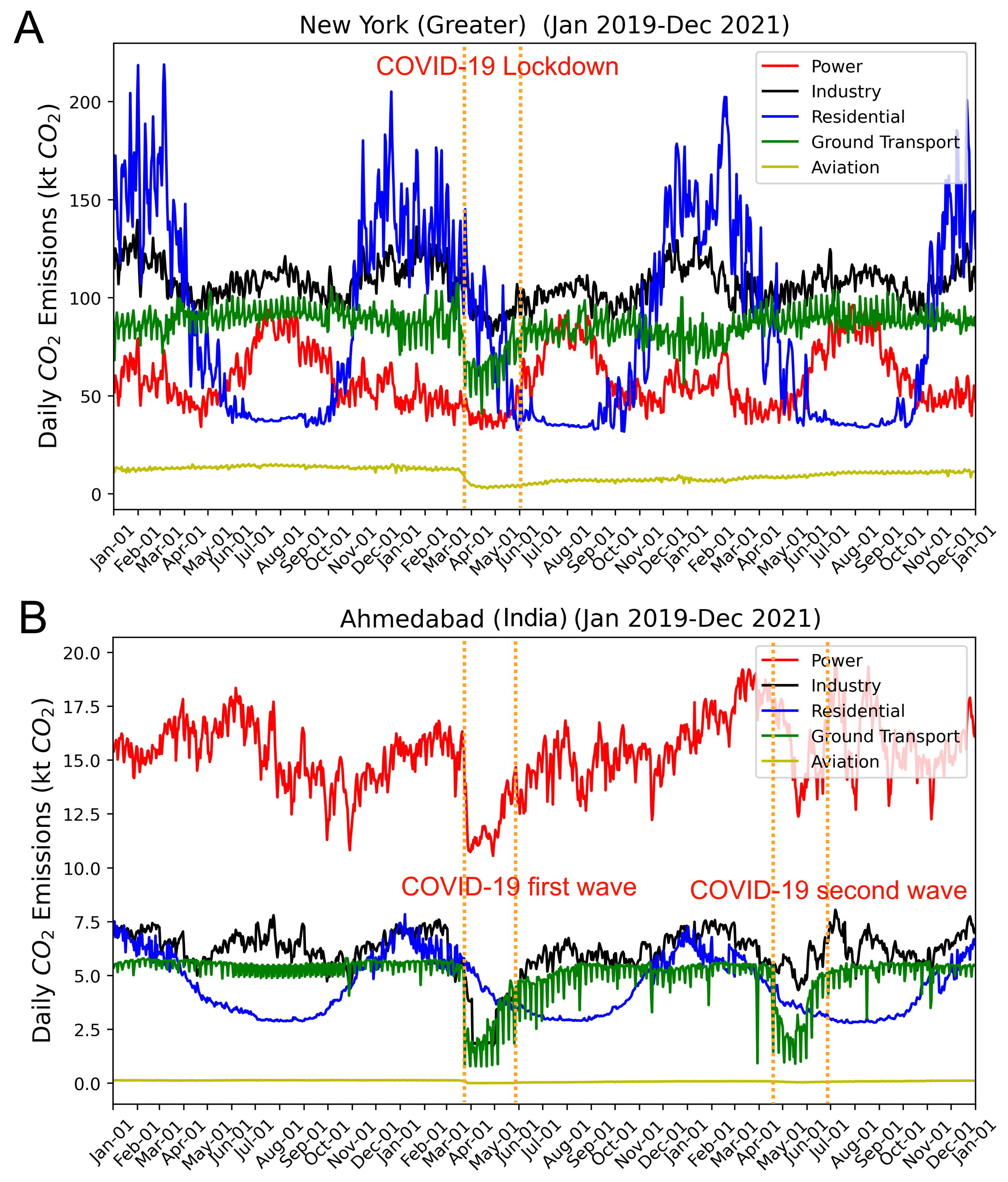}
\caption{Daily city-level CO$_2$ emissions show the impact of COVID-19 for (A) Greater New York in the U.S. and (B) Ahmedabad in India. The emissions from ground transportation and aviation decreased significantly during the lockdown period (between the dashed lines) in spring 2020, and also during the second wave between March 2021 to June 2021 in India.}
\label{fig:fig5}
\end{figure}

Daily CO$_2$ emission variations from a city reveal its geographic and socio-economic characteristics. Figure \ref{fig:fig4} shows the sectoral breakdown of daily CO$_2$ emissions for some major cities in different regions of the world, including East Asia, Middle East, Southeast Asia, West Europe, East Europe, Oceania, South America, North America, and Africa. As an example of the geographic influence on the emissions, cities in the Southern Hemisphere, such as Sydney in Australia, and Cape Town in Africa, exhibit higher emissions in the residential sector during the southern winter (northern summer) due to the increase in heating demand. Daily emissions also reveal certain events such as holidays and the COVID-19 outbreak. The emissions from the power sector show a surge in summer for many cities, which is likely due to the increased power consumption for cooling. As an example, we show the impact of COVID-19 on city emissions for Greater New York in the U.S. and Ahmedabad in India (Fig. \ref{fig:fig5}), note the significant drop in emissions for the ground transport sector in spring 2020 (as indicated by the dashed lines) during the lockdown period and during the 2021 second wave of COVID-19 pandemic in India.

\section*{Technical Validation}
\begin{figure}[ht]
\centering
\includegraphics[width=12cm]{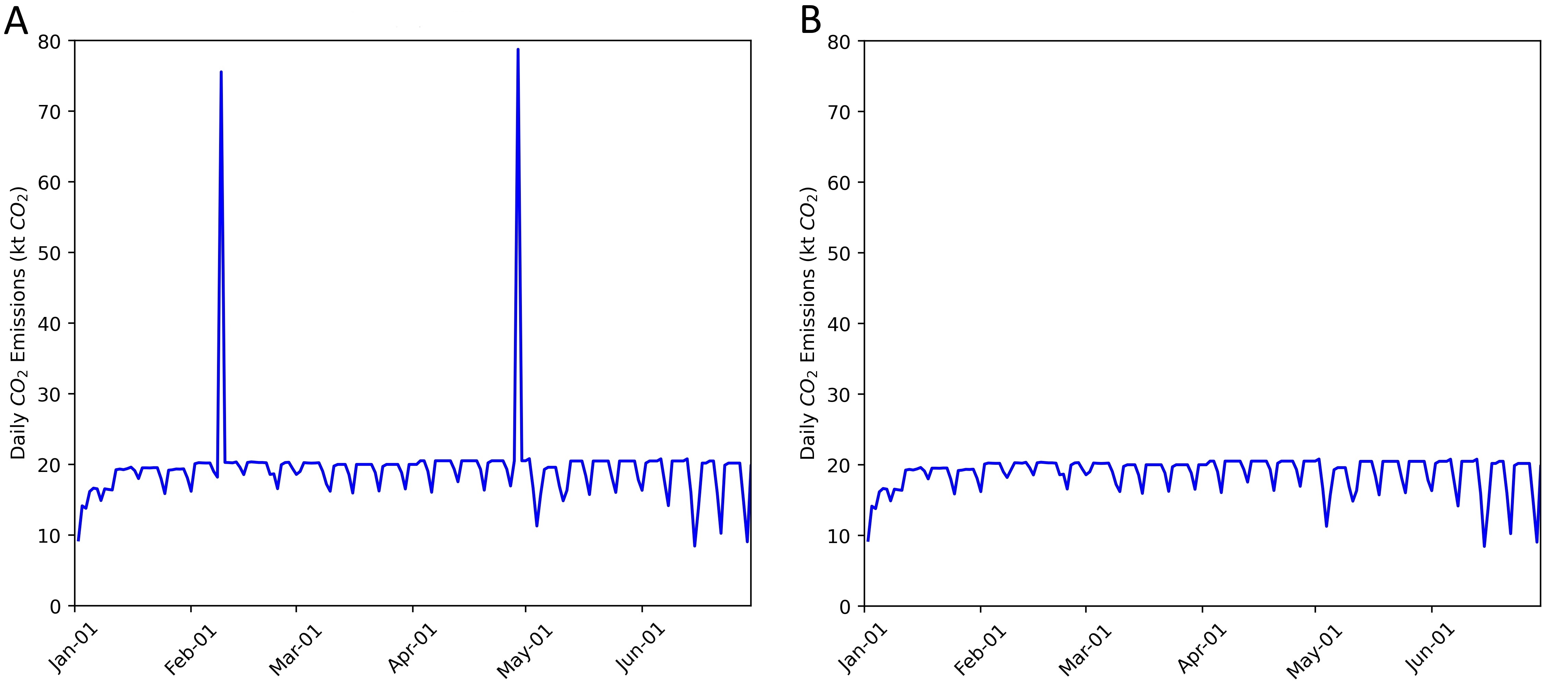}
\caption{Examples of outlier identification and correction for the ground transport data. (A) Two outliers clearly fall out of the typical range of weekday-weekend variation before the correction. (B) Outliers removed after the correction.}
\label{fig:fig6}
\end{figure}

The quality of this dataset is evaluated by comparing it against existing datasets (Table \ref{tab:tab5}). We also performed uncertainty analysis for our data and for each sector based on a synthesis analysis of input data uncertainties and the methodology used. Significant outliers were identified and corrected as shown in the examples of Figure \ref{fig:fig6}. The outlier occurrence rate for this dataset is 0.012\%.

\subsection*{Validation Against Other Datasets}
\begin{figure}[ht]
\centering
\includegraphics[width=15cm]{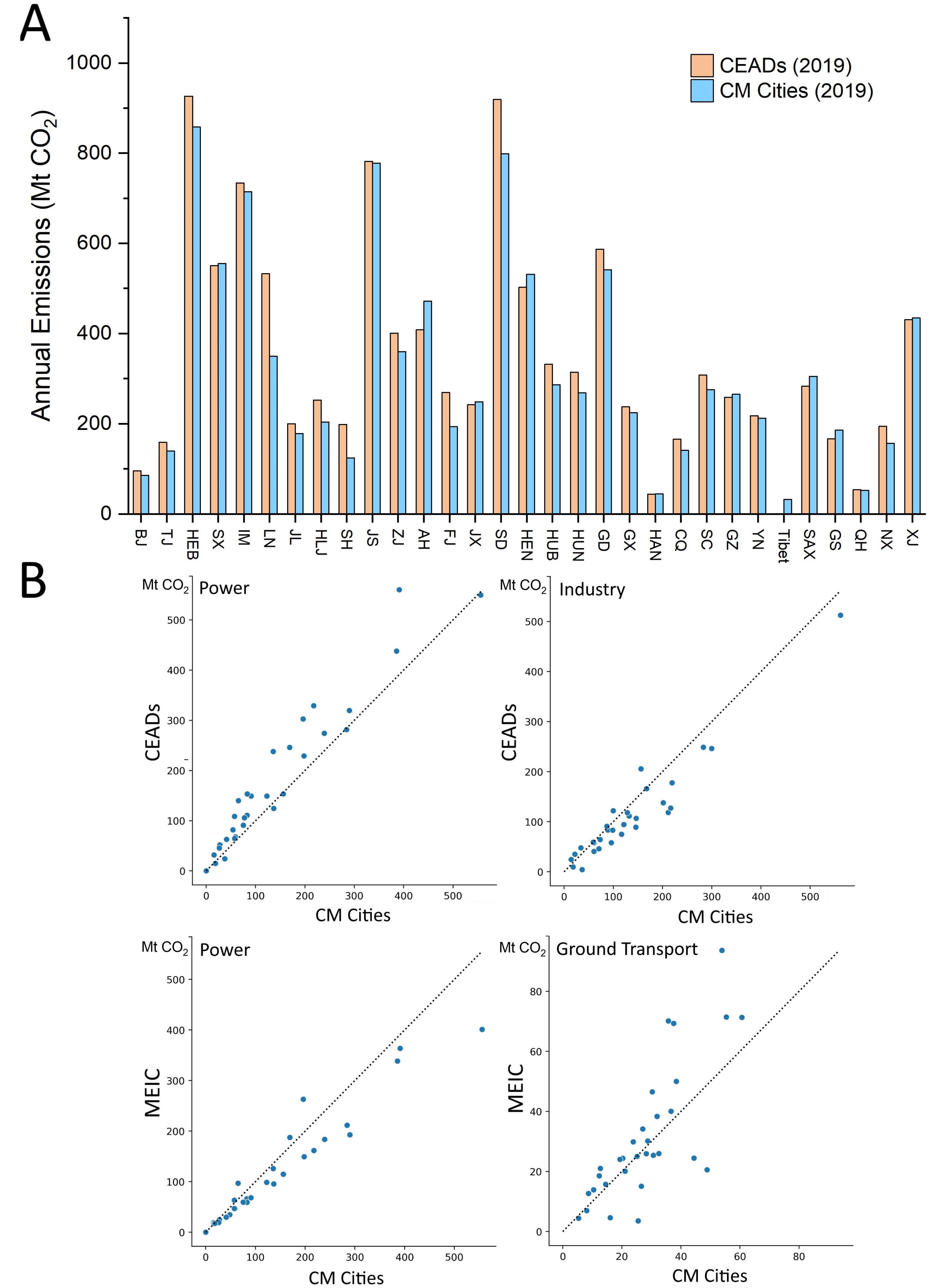}
\caption{Dataset comparison for cities in China. (A) Comparison of the sum of all prefecture-level cities within each Chinese province (including municipalities and autonomous regions) against the CEADs provincial datasets for year 2019. Note that the sum of all prefecture-level cities within a province equals the total area of that province in China (B) Sectoral comparison between CM Cities, CEADs and MEIC for sectors that have similar coverages.}
\label{fig:fig7}
\end{figure}

Multiple datasets are used to validate our results, including 1. City inventories from the CDP, 2. Vulcan dataset for US counties, 3. CEADs and MEIC dataset for China, and 4. individual reports released by city governments. For cities in China, we validated our dataset by comparing it with the CEADs and MEIC datasets. CEADs provides annual provincial emission inventories for China for 2019, and we validated the data for each province by summing up emissions from all prefecture-level cities in each province (for China, the GADM level-2 is exactly the area of prefecture-level cities, and the sum of all prefecture-level cities within a province equals the total area of that province). Figure \ref{fig:fig7} depicts the comparison results for all the Chinese provinces including municipalities and most autonomous regions. The result indicates a good agreement between the CEADs and CM-Cities, with less than 10\% difference in annual emissions for most of the provinces. Statistics (Table \ref{tab:tab5}) indicate that the coefficient of determination ($R^2$) values between CEADs and CM-Cities are 0.96, 0.76, and 0.92 for total, power, and industry sectors, respectively, and the corresponding mean relative difference ($Rd$) are 11\%, 30\%, and 28\%. $R^2$ values between MEIC and CM-Cities are 0.93 and 0.62 for the power sector and ground transport sector, respectively, and the corresponding $Rd$ are 21\% and 31\%. Other sectors were not compared due to the large differences in sector definition and coverage. The mean relative differences are all within the uncertainty ranges, which indicates a relatively high accuracy for Chinese cities.

\begin{figure}[ht]
\centering
\includegraphics[width=15cm]{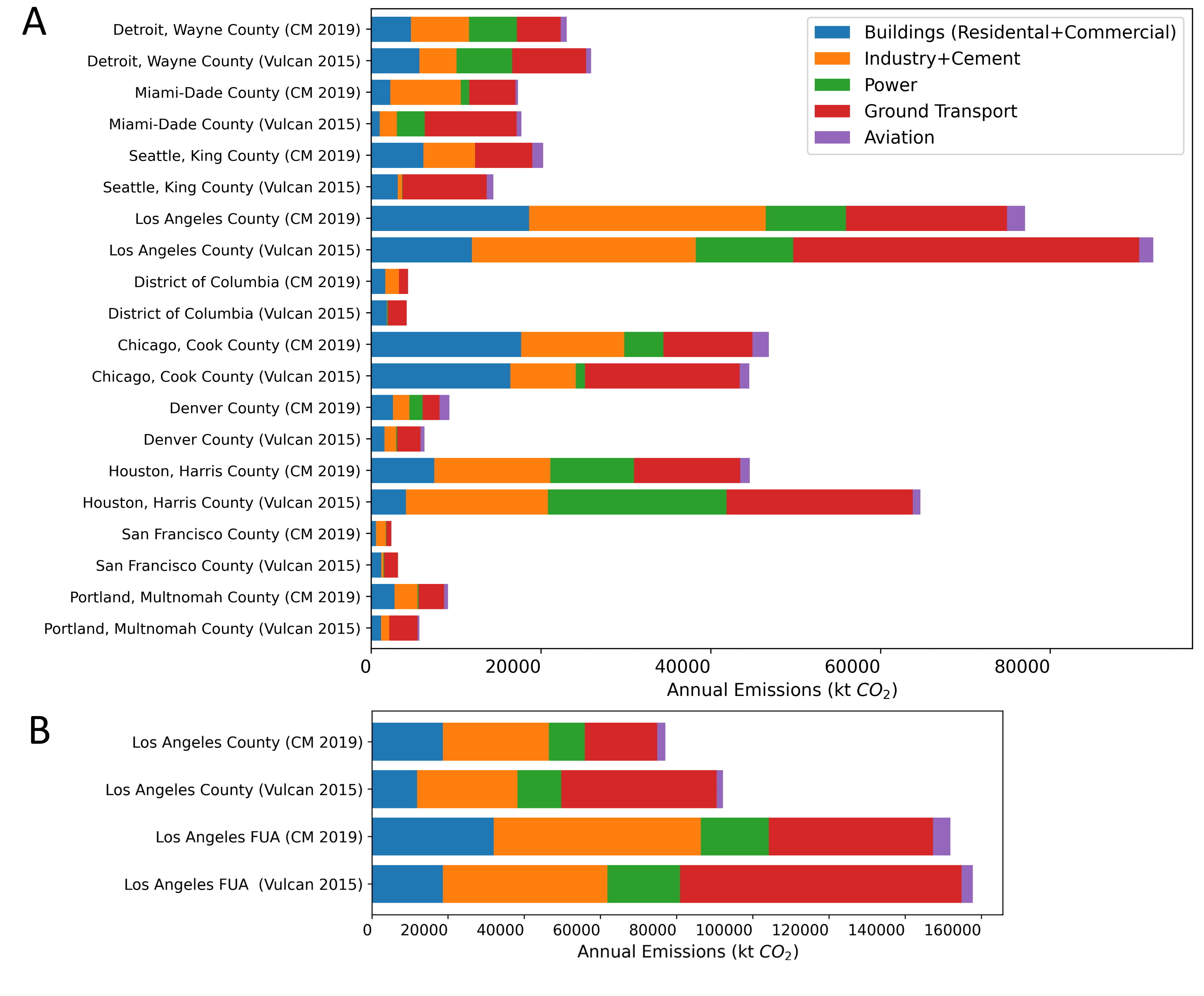}
\caption{Comparison of sectoral emissions between Vulcan dataset and CM-Cities for (A) selected counties in the United States, and (B) county-level and FUA-level comparison. The year of accounting is 2015 for Vulcan inventories and 2019 for CM-Cities, which could partially explain the differences.}
\label{fig:fig8}
\end{figure}

\begin{figure}[ht]
\centering
\includegraphics[width=\linewidth]{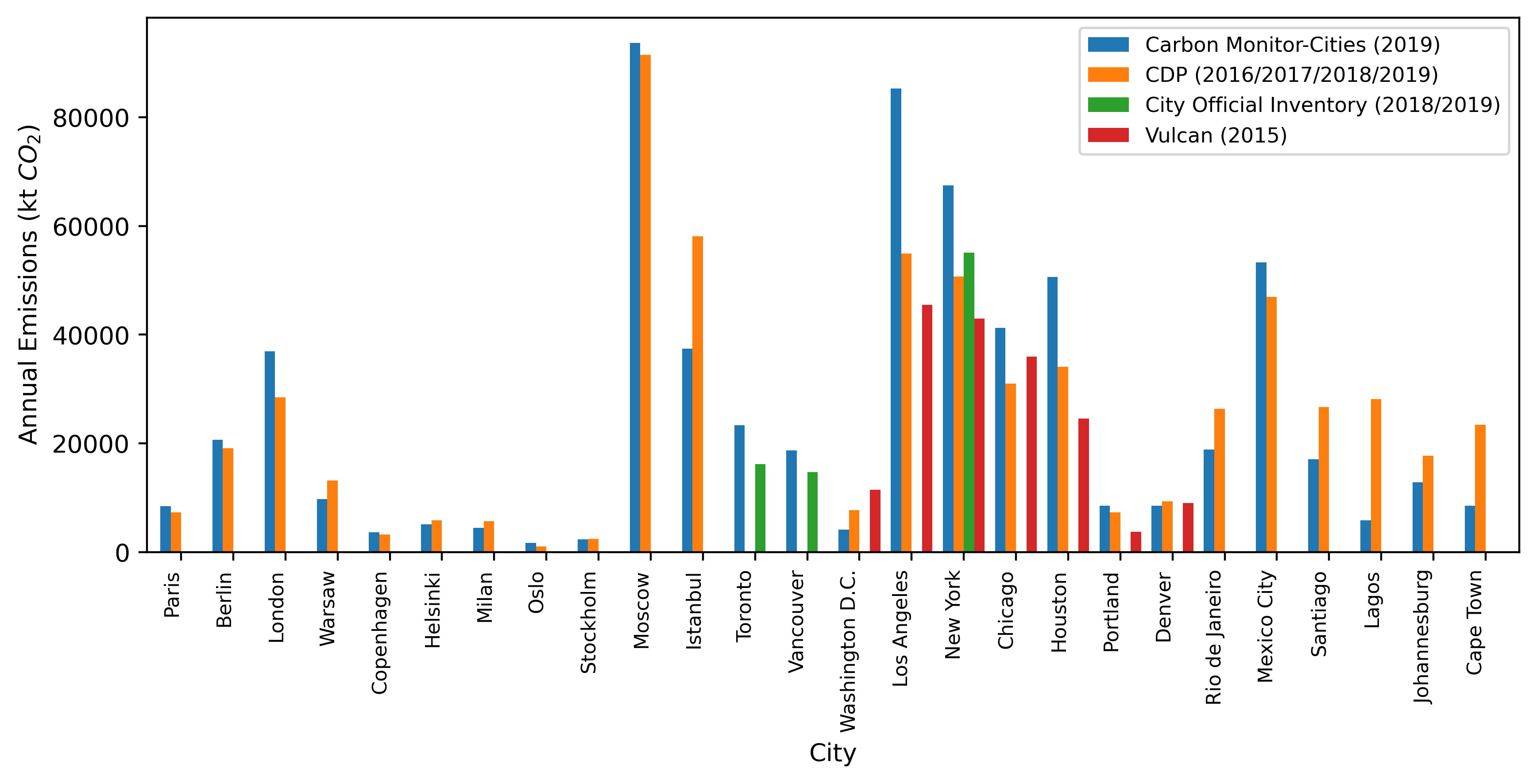}
\caption{City annual total emission comparisons between CM-Cities, CDP, Vulcan and some other city self-reported inventories. Magnitudes represent total emissions from each dataset. The area of accounting is adjusted to be as consistent as possible across datasets.}
\label{fig:fig9}
\end{figure}

For cities in other countries, many datasets do not have recent (2019 or later) inventories, for example, the latest Vulcan dataset provides emissions in 2015 for United States counties, and the latest CDP inventories may range from 2010 to 2021 depending on reporting status of each city. For the completeness of the validation, we adjusted the area of accounting for CM-Cities to be as consistent as possible with these datasets and compared city inventories from all available data sources regardless of the time of accounting. Note that GADM level-2 in the United States represents exactly the area of counties, so our GADM results were used for comparison with Vulcan county-level inventories. Figure \ref{fig:fig8} and Figure \ref{fig:fig9} show examples of the annual total emission comparisons between CM-Cities and these datasets.

Comparison between Vulcan (2015) and CM-Cities (2019) covers top 50 counties with the highest emissions in the United States. We used the coefficient of determination ($R^2$) and mean relative difference ($Rd$) to evaluate the comparison results. $R^2$ values are 0.82, 0.60, 0.58, 0.82, 0.90, and 0.69  for total, power, industry, residential (and commercial buildings), ground transport, and aviation sectors, respectively, and the corresponding $Rd$ values are 26\%, 114\%, 67\%, 35\%, 41\%, and 58\%, respectively (Table \ref{tab:tab5}). The differences are mainly due to 1. the difference in the year of accounting, as the earliest estimates of CM-Cities for the year 2019 is compared to the latest Vulcan for the year 2015, multiple factors that govern emissions could have changed during the period, 2. the different accounting methods, as CM-Cities uses a territorial downscaling approach, while Vulcan uses a consumption-based bottom-up accounting approach, and 3. the differences in sector coverage definitions and source data (Tables \ref{tab:tab1}, \ref{tab:tab2}, \ref{tab:tab5}), which partially explains why the total emission comparison show a better good agreement than the sectoral comparisons.
 
Direct comparisons with CDP were difficult due to several reasons: 1. CDP inventories are city self-reported data, which were typically estimated using different methods. 2. Most cities follow the GPC protocol and report in scopes rather than in sectors, therefore, we only compared total emissions. 3. CDP has not independently calculated the uncertainty range for these self-reported inventories, and self-reported uncertainties are expected to be variable. For example, 45\% of cities reported “high confidence” in their emissions data for 2021, 35\% reported “medium confidence”, and 3\% reported “low confidence”. 4. Spatial coverage is unclear, as the definition of a 'city' can vary across different countries, some cities report based on administrative areas, but others include adjacent areas, but no shapefiles or raster maps were provided to clarify the exact city boundary or area of accounting. Nonetheless, we performed comparisons for 24 large cities in different regions with available CDP inventories, the total emission comparisons for these cities show an agreement with $R^2$ = 0.74 and $Rd$ = 31\% (Fig. \ref{fig:fig9}, Table \ref{tab:tab5}).

\subsection*{Uncertainty Analysis}
The uncertainties in this dataset have two sources: 1. The uncertainties inherited from Carbon Monitor and GRACED. 2. The uncertainty introduced by the spatial downscaling process. The uncertainty analysis was conducted based on the 2006 IPCC Guidelines for National Greenhouse Gas Inventories. For the power sector, uncertainty mainly comes from the emission factor and the variability of energy mix for power generation, the 1-sigma uncertainty of power emission from fossil fuel is estimated as ±10.0\%. For the industry sector, monthly production data is the main source of uncertainty, especially the production in China, which accounts for more than 60\% of world total industrial CO$_2$ emissions. Monte Carlo simulations were used to determine the confidence interval based on regression models between estimated monthly emissions and officially reported emissions. The 1-sigma uncertainty for the industry sector is estimated as ±36.0\%. For the ground transport, uncertainty is estimated by applying the regression between the TomTom congestion index and traffic flux to other cities (other than Paris). The 1-sigma uncertainty for the ground transport sector is estimated as ±9.3\%. For the residential sector, the uncertainty is calculated based on comparisons between estimated emissions and consumption-based accounting results for several countries in Europe. The 1-sigma uncertainty for the residential sector is estimated as ±40.0\%. For the aviation sector, 1-sigma uncertainty is estimated as ±10.2\%. These uncertainty estimates follow the methods used by Carbon Monitor \cite{Liu2020}.

Spatial downscaling introduces additional uncertainty because of the rasterization of city areas. Spatial computations are based on raster (gridded) files, but most cities and urban areas have irregular-shaped boundaries that are not fully overlapped with gridded cells. Area discrepancies are found along all city boundaries, and smaller cities typically suffer from higher levels of dissimilarities because few grid cells account for a large portion of the total urban area. We computed the area discrepancies for all cities in the dataset, and found that 44.53\% of cities show an area difference of 0\%-5\% and the count decreases as the discrepancy percentage gets higher. The mean area discrepancy for all cities is 13.55\%. We then estimated the overall uncertainty by first applying the error propagation equation provided by IPCC\cite{IPCC2006}, and then combining the uncertainties of all sectors and the city area uncertainty: 

\begin{equation}
U_{overall}=\sqrt{\frac{\sum{(U_{s}\alpha_{s})}}{{\sum\alpha_{s}}^2}}+U_{a},
\end{equation}

\noindent where $U_{s}$ and $\alpha_{s}$ are the percentage and quantity (daily mean emissions) of the uncertainty for sector $s$, respectively, and $U_{a}$ is the city area uncertainty. Finally, the overall uncertainty range of CM-Cities is estimated as ±21.66\%.

\section*{Usage Notes}

The generated datasets are available from \hyperlink{https://doi.org/10.6084/m9.figshare.19425665.v1}{https://doi.org/10.6084/m9.figshare.19425665.v1}.The main data file has more than one million lines of data, which will take a long time to load in Excel. We recommend loading the data with a script that can handle large datasets. We have provided an example of Python code to help users read in and plot emissions for any city in the dataset (\hyperlink{https://github.com/dh107/Carbon-Monitor-Cities/}{https://github.com/dh107/Carbon-Monitor-Cities/}). Users should also note that the unit of emissions in this dataset is $kt CO_2$. Filename indicates whether the data is based on administrative areas (GADM) or functional urban areas (default). The next update to this dataset is scheduled for May 2022, which will update the dataset to Feb 28, 2022.

\section*{Code availability}
Python code for producing, reading and plotting data for any city in the dataset is provided  at \hyperlink{https://github.com/dh107/Carbon-Monitor-Cities/}{https://github.com/dh107/Carbon-Monitor-Cities/}.

\bibliography{sample}

\section*{Acknowledgements}
Authors acknowledges support from the CDP Worldwide.

\section*{Author contributions statement}
D.H. and Z.L. designed the research. D.H., X.H. and X.D. conducted the data processing and wrote the manuscript. D.H., P.C. Z.D. and Z.L. designed the methods, and all authors contributed to data collection, discussion and analysis.

\section*{Competing interests}
The authors declare no competing interests.

\section*{Figures \& Tables}

\begin{table}[ht]
\centering
\begin{tabular}{p{0.15\linewidth} p{0.45\linewidth} p{0.3\linewidth}}
\hline
Country/Region & Data source & Description \\
\hline
China & National Grid Daily Electric Load & Daily thermal production \\
India & Power System Operation Corporation Limited (https://posoco.in/reports/daily-reports/) & Daily thermal production (from coal, lignite, gas, naphtha and diesel) \\
United States & Energy Information Administration’s (EIA) Hourly Electric Grid Monitor (https://www.eia.gov/beta/electricity/gridmonitor/) & Hourly thermal production (from coal, petroleum, and natural gas) \\
EU27 & ENTSO-E Transparency platform (https://transparency.entsoe.eu/dashboard/show) & Hourly thermal production \\
United Kingdom & Balancing Mechanism Reporting Service (BMRS) (https://www.bmreports.com/) & Hourly power generation \\
Russia & United Power System of Russia (http://www.so-ups.ru/index.php) & Total hourly generation \\
Japan & Organization for Cross-regional Coordination of Transmission Operators (OCCTO) (https://www.occto.or.jp/en/) & Hourly thermal generation \\
Brazil & Operator of the National Electricity System (http://www.ons.org.br/Paginas/) & Hourly thermal production \\
\hline
\end{tabular}
\caption{\label{tab:tab1}Data sources for the power sector.}
\end{table}

\begin{table}[ht]
\centering
\begin{tabular}{p{0.12\linewidth} p{0.11\linewidth} p{0.35\linewidth} p{0.32\linewidth}}
\hline
Country/Region & Sector & Data & Data Source \\
\hline
China & Steel industry & Crude steel production & World Steel Association (https://www.worldsteel.org/) \\
 & Cement industry & Cement and clinker production & National Bureau of Statistics (http://www.stats.gov.cn/english/)\\
 & Chemical industry & Sulfuric acid, caustic soda, soda ash, ethylene, chemical fertilizer, chemical pesticide, primary plastic and synthetic rubber & National Bureau of Statistics (http://www.stats.gov.cn/english/)\\
 & Other industry & Crude iron ore, phosphate ore, salt, feed, refined edible vegetable oil, fresh and frozen meat, milk products, liquor, soft drinks, wine, beer, tobaccos, yarn, cloth, silk and woven fabric, machine-made paper and paperboards, plain glass, ten kinds of nonferrous metals, refined copper, lead, zinc, electrolyzed aluminum, industrial boilers, metal smelting equipment, and cement equipment & National Bureau of Statistics (http://www.stats.gov.cn/english/)\\
India & - & Industrial Production Index (IPI) & Ministry of Statistics and Programme Implementation (http://www.mospi.nic.in) Trading Economics (https://tradingeconomics.com) \\
United States & - & Industrial Production Index (IPI) & Federal Reserve Board (https://www.federalreserve.gov) \\
EU27 and UK & - & Industrial Production Index (IPI) & Eurostat (https://ec.europa.eu/eurostat/home) Trading Economics (https://tradingeconomics.com) \\
Russia & - & Industrial Production Index (IPI) & Federal State Statistics Service (https://eng.gks.ru) \\
Japan & - & Industrial Production Index (IPI) & Ministry of Economy, Trade and Industry (https://www.meti.go.jp) \\
Brazil & - & Industrial Production Index (IPI) & Brazilian Institute of Geography and Statistics (https://www.ibge.gov.br/en/institutional/the-ibge.htm) \\
\hline
\end{tabular}
\caption{\label{tab:tab2}Data sources for industrial production.}
\end{table}

\begin{table}[ht]
\centering
\begin{tabular}{p{0.15\linewidth} p{0.5\linewidth} p{0.25\linewidth}}
\hline
Data & Data Description & Resolution  \\
\hline
Global Carbon Grid (GID) v1.0  & Global Grid in the Global Infrastructure Emission Database. Including power, industry, residential, transport, shipping, and aviation sectors with high data quality in spatial fine-grained maps (http://gidmodel.org) & 0.1$^{\circ}$ × 0.1$^{\circ}$ global, Annual \\
Emission Database for Global Atmospheric Research (EDGARv5.0) & EDGAR v5.0 FT2019, covers major fossil CO$_2$ sources globally, with monthly emissions provided per main source category (https://edgar.jrc.ec.europa.eu)  & 0.1$^{\circ}$ × 0.1$^{\circ}$ global, Monthly \\
TROPOMI NO$_2$ Retrievals & NO$_2$  thermal chemical vapor deposition retrievals acquired by the TROPOspheric Monitoring Instrument (TROPOMI) on board the Sentinel‐5 Precursor satellite, launched in 2017  & 0.1$^{\circ}$ × 0.1$^{\circ}$ global, Daily \\
\hline
\end{tabular}
\caption{\label{tab:tab3}List of gridded data used for producing Global Gridded Daily CO$_2$ Emissions Dataset (GRACED).}
\end{table}

\begin{table}[ht]
\centering
\begin{tabular}{p{0.1\linewidth} p{0.6\linewidth}}
\hline
Column & Description \\
\hline
City & Name of the city \\
Country & Country where the city is located. The following countries are covered in this dataset: Argentina, Australia, Austria, Bangladesh, Belgium, Brazil, Canada, Chile ,China, Colombia, Denmark, Egypt, Finland, France, Germany, Greece, Hungary, India, Indonesia, Iran, Italy, Japan, Korea, Malaysia, Mexico, Myanmar, Netherlands, Nigeria, Norway, Pakistan, Peru, Philippines, Poland, Portugal, Russia, Singapore, South Africa, Spain, Sweden, Switzerland, Thailand, Turkey, United Arab Emirates, United Kingdom, United States, Vietnam\\
Date & Date (YYYY-MM-DD) on which the emissions were estimated. Currently, the dataset provides emissions from 2019-01-01 to 2021-12-31 \\
Sector & Sector for which the emissions were estimated, including power, industry, residential, ground transport, aviation\\
Value & Magnitude of daily emissions with a unit of $kt CO_2$\\
Timestamp & Unix timestamp at 00:00:00 (GMT+0000) on each day for scientific visualization\\
\hline
\end{tabular}
\caption{\label{tab:tab4}Data attributes.}
\end{table}

\begin{table}[ht]
\centering
\begin{tabular}{p{0.18\linewidth} p{0.13\linewidth} p{0.13\linewidth} p{0.1\linewidth} p{0.2\linewidth} p{0.15\linewidth}}
\hline
Dataset & CM Cities & CEADs & MEIC & CDP & Vulcan \\
\hline
Spatial coverage & Global cities & China national, provincial, prefectural & China national, provincial & Global cities & U.S. counties \\
Temporal coverage & 2019-2021 & 1997-2019 & 2000-2017 & 2010-2021 & 2010-2015 \\
Temporal resolution & Daily & Monthly & Annual & Annual & Annual, hourly \\
Protocol & - & - & - & Various & - \\
Overall uncertainty & ±21.7\% & -15\% to 30\% & ±15\% & All data is self-reported, CDP does not assess the uncertainty & Sectoral uncertainties provided below \\
Area definition & GADM, FUA & N/A & Population density, GDP & Mostly city administrative, some include adjacent areas & Administrative county area \\
Total emissions comparison (with CM Cities) & - & $R^2$=0.96, $Rd$=11\%, $n$=30 & - & $R^2$=0.74, $Rd$=31\%, $n$=24 & $R^2$=0.82, $Rd$=26\%, $n$=50 \\
Power sector method & Daily power generation downscaling. $\Delta$= ±10\% &  Energy consumption for production and supply of electric power, steam and hot water & Unit-level power generation. $\Delta$= -15\% to 16\% & City report (scope 1-3 for relevant GPC stationary energy subsectors, including residential and commercial buildings, industry, agriculture, forestry and fishing) & CAMD, DOE/ EIA fuel, EPA NEI point electricity production. $\Delta$= ±13\% \\
Power comparison (with CM Cities) & - & $R^2$=0.76, $Rd$=30\%, $n$=30 & $R^2$=0.93, $Rd$=21\%, $n$=30 & - & $R^2$=0.60, $Rd$=114\%, $n$=50 \\
Industry sector method & Industrial production index downscaling. $\Delta$= ±36\% & Energy consumption for individual manufacturing sectors & - & City report (direct scope 1 emissions from industrial processes and product use) & EPA NEI industrial point sources. $\Delta$= ±12.8\%\\
Industry comparison (with CM Cities) & - & $R^2$=0.92, $Rd$=28\%, $n$=30 & - & - & $R^2$=0.58, $Rd$=67\%, $n$=50 \\
Residential sector method & HDD. $\Delta$= ±40\% & - & - & City report (scope 1-3 for relevant GPC stationary energy subsectors, including residential and commercial buildings) & EPA NEI residential and commercial nonpoint buildings. $\Delta$= ±12.8\% \\
Residential comparison (with CM Cities) & - & - & - & - & $R^2$=0.82, $Rd$=35\%, $n$=50 \\
Ground transport sector method & TomTom congestion index. $\Delta$= ±9.3\%  & - & Vehicle ownership statistics and digital road map &  City report (scope 1-3 for GPC transportation subsectors, including on-road, railways, waterborne navigation, aviation, and off-road) & EMFAC, EPA NEI onroad. $\Delta$= ±14.2\% \\
Ground transport comparison (with CM Cities) & - & - & $R^2$=0.62, $Rd$=31\%, $n$=30 & - & $R^2$=0.90, $Rd$=41\%, $n$=50 \\
Aviation sector method & Flightradar24 flight data. $\Delta$= ±10.2\% & - & - &  City report aviation under transportation sector & EPA NEI point airport. $\Delta$= ±7.8\% \\
Aviation comparison (with CM Cities) & - & - & - & - & $R^2$=0.69, $Rd$=58\%, $n$=50 \\
References & - & \cite{Shan2020} & \cite{Zheng2014,Liu2015} & - & \cite{Gurney2020} \\
\hline
\end{tabular}
\caption{\label{tab:tab5}Summary of city emission datasets ($\Delta$ is uncertainty) and comparison statistics including coefficient of determination ($R^2$), mean relative difference ($Rd$), and sample size ($n$) when compared with CM Cities.}
\end{table}

\end{document}